# Construction of Yemilab

K. S. Park[1], Y. D. Kim[1,2], K. M. Bang[1], H. K Park[3], M. H. Lee[1,2], J. H. Jang[1], J. H. Kim[1], J. So[1], S. H. Kim[1], S. B. Kim[1]

[1]Center for Underground Physics, Institute for Basic Science (IBS), Daejeon 34126, Republic of Korea

[2]IBS School, University of Science and Technology (UST), Daejeon, 34113, Republic of Korea

[3]Department of Accelerator Science, Korea University, Sejong 30019, Republic of Korea

**\* Correspondence:** K.S. Park
heppark@ibs.re.kr

## Abstract

The Center for Underground Physics of the Institute for Basic Science (IBS) in Korea has been planning the construction of a deep underground laboratory since 2013 to search for extremely rare interactions such as dark matter and neutrinos. In September 2022, a new underground laboratory, Yemilab, was finally completed in Jeongseon, Gangwon Province, with a depth of 1,000 m and an exclusive experimental area spanning 3,000 m$^3$. The tunnel is encased in limestone and accommodates 17 independent experimental spaces. Over two years, from 2023 to 2024, the Yangyang Underground Laboratory facilities will be relocated to Yemilab. Preparations are underway for the AMoRE-II, a neutrinoless double beta decay experiment, scheduled to begin in Q2 2024 at Yemilab. Additionally, Yemilab includes a cylindrical pit with a volume of approximately 6,300 m$^3$, designed as a multipurpose laboratory for next-generation experiments involving neutrinos, dark matter, and related research. This article provides a focused overview of the construction and structure of Yemilab.

## 1    Introduction

The Institute for Basic Science (IBS) is a Korean national research institute established in 2011 to conduct basic scientific research. As of 2023, IBS includes 31 independent centers, each focusing on different research themes. The Center for Underground Physics (CUP) is one of the centers and, as indicated by its name, is a research center dedicated to astroparticle physics, conducting primary experiments underground. CUP started in 2013, and approximately 70 researchers are engaged in research activities to measure extremely rare interactions, such as dark matter and neutrinos. Since its inception, CUP planned to construct a new underground laboratory to expand the existing Yangyang Underground Laboratory (Y2L), which has a limited experimental space of ~200 m$^2$ at a depth of ~700 m. In September 2022, a new underground experimental facility with a depth of 1,000 m (2,500 m w.e.) and an exclusive experimental area of ~3,000 m$^2$ was finally completed in Jeongseon-gun, Gangwon province, named the Yemilab. All tunnels in the experimental area are predominantly surrounded by limestone, providing more than ten independent experimental sections.

## 2    Site Selection

The implementation of extremely low background radiation, a crucial aspect in underground experiment research, can only be achieved by thoroughly understanding and addressing various complex factors. One of the most essential factors is the shielding of muons. Several locations were considered for the site of the new underground experimental facility. Firstly, existing closed or operational mines were investigated. Another option involved excavating tunnels on high mountains to secure deep underground spaces. Candidate sites were selected, considering tunnel length and lab depth. After a comprehensive investigation spanning about three years, the options narrowed down to two locations in Gangwon province: Mt. Duta (peak height 1,357 m) located in Samcheok-city and Mt. Yemi (peak height 989 m) encompassing SM Handuk Iron Mine Co., Ltd (Handuk) in Jeongseon-gun. Both sites allowed for the construction of underground experimental facilities at depths exceeding 1,000 m. Subsequently, after conducting a feasibility study that considered geological conditions, excavation and construction costs, permission issues, and more, Mt. Yemi was confirmed as the final site for the new underground research facility.

The primary reason for choosing Mt. Yemi was its high muon shielding effect compared to the construction cost and a relatively simple permission process. Firstly, when considering construction costs, it was estimated that the construction cost for Mt. Yemi was nearly half of that for Mt. Duta. It was attributed to the 627 m long shaft owned by Handuk near the top of Mt. Yemi. By utilizing this shaft, the access tunnel length to be excavated was only about half compared to Mt. Duta, resulting in significant cost savings. Secondly, permission issues related to construction were a concern. For Mt. Duta, excavating a new tunnel was subject to multiple regulations concerning natural preservation, necessitating careful consideration of numerous procedures and restrictions. On the other hand, as the Mt. Yemi site employed the existing operational Handuk, construction could proceed without any environmental damages, and there were no significant legal restrictions or permission issues. Lastly, the third reason was muon shielding performance. The possible depths of the underground experimental facilities at Mt. Duta and Mt. Yemi were 1,400 and 1,000 m, respectively, resulting in a difference. When calculating muon rates using the topographies of both mountains, the muon flux at the depth just under the summit of the mountains was reduced to approximately $1.8 \times 10^{-6}$ at Mt. Duta and $3 \times 10^{-6}$ at Mt Yemi compared with the muon rate at sea level. The difference between the two is less than a factor of 2. Despite the depth difference, which might suggest a factor of 5-6, the influence of omnidirectional coverage reduces the expected factor. It could be attributed to better omnidirectional coverage in Yemi compared to Duta. While Mt. Duta had about 40% higher shielding capability compared to Mt. Yemi, both have over five times the shielding efficiency of the Y2L. Thus, they deemed suitable as potential new sites. After considering all these factors, it was decided in mid-2015 to construct a new deep and spacious underground research facility at Mt. Yemi at a depth of 1,000 m. Immediately, construction plans were formulated to secure the budget. The following year, after passing the National Facility Equipment Committee review, construction of the Yemilab began in September 2017, starting with the installation of an elevator for personnel access called man-cage. The construction took place over a five-year period until 2022.

## 2.1 Geology Overview

Handeok is located in Jodong-ri, Sindong-eup, Jeongseon-gun, Gangwon province, South Korea. It is situated west of the Mt. Taebaek mineralized zone. The surrounding terrain is characterized by east-west oriented ridges such as Mts. Yemi (989.2 m), Baejae (740 m), and Jilun (1171.8 m). The southern slope, delineated by these ridges, is steep, while the northern slope features a gentler rugged terrain. The variation in slope gradient is attributed to the distribution of rock types, with sandstone and shale predominantly found on the northern side, whereas limestone predominates on the southern side. In the south part of the area, the geological composition includes the upper layer of the Makdong limestone formation, part of the Paleozoic Joseon Supergroup, overlain by the Hongjom series from the Pyeongan Supergroup, and intrusions of quartz diorite/quartz diorite gneiss of unknown age. The upper part of the Makdong limestone formation is overlain by a Jikunsan shale formation, characterized by its color ranging from black to dark gray.

## 3   Construction Timeline

Since 2013, CUP has been exploring potential sites to construct a new underground research facility. In early 2016, construction proceeded in stages at the final location at Mt. Yemi (Figure 1). By the end of 2016, funding for the first construction phase was secured. The first phase involved the installation of the man-cage and primary tunnel excavation. The cage installation was completed in December 2018, and the first construction phase was concluded with the tunnel construction in August 2020. In early 2021, funding for the second phase of construction was secured, and work for the second phase resumed at the end of 2021. The second phase included excavation work for a pit to accommodate a 6200 m$^3$ Large Scintillation Counter (LSC) and the installation of electrical, ventilation, communication, and fire safety facilities underground. The second construction phase was completed in August 2022, establishing the basic framework of Yemilab. With the completion of the fundamental structure of Yemilab, interior enhancements took place in 2023 (Figure 2). They involved the installation of equipment such as cranes, vehicle washing systems, dustproof doors, and coating for floors and walls. Finally, Yemilab acquired the necessary infrastructure for the beginning of experiments. Starting in September 2023, the relocation of various facilities from Yangyang to Yemilab began. By the end of 2024, once all the relocations are completed, each experiment will enter a new phase at Yemilab.

## 4   Tunnel

In February 2019, the first blast for tunnel excavation started. The Yemilab tunnel starts approximately 100 m away from the shaft. The tunneling progressed towards the peak of Mt. Yemi. While some sections of the rock at the excavation site exhibited quartzite and traces of iron ore, most of the experimental section was composed of limestone.

### 4.1   Design

Yemilab consists of seventeen dedicated and independent experimental rooms. The basic structure is designed with two exits for experimental rooms exceeding a length of over 20 meters to ensure that researchers can safely and swiftly escape in the event of a fire. Therefore, it inherently features a ladder-type design. In this structure, creating clean rooms according to users' preferences is more efficient than the structure of a large hall.

The starting point of the Yemilab tunnel is situated at an elevation of -35 m in the Handuk area, with a depth of 604 m. Therefore, the Yemilab tunnel does not extend to connect with any other public roads. From the tunnel front, an entry tunnel was excavated with a standard cross-section of 5 m (width) by 5 m (height), featuring a downward slope of 12% and a length of 782 m. This slope was intentionally designed to achieve a greater depth, aiming for a descent of approximately 100 m at the end of the tunnel. Turning shelters were established every 90 m to facilitate the rotation of large construction vehicles. There are six turning shelters in the access tunnel, four of which can be repurposed as experimental spaces.  A horizontal tunnel for experiments was excavated from the end of the access tunnel. In the experimental area, 23 independent spaces were excavated, with 16 dedicated to experiments or housing equipment directly supporting experiments and the remainder serving operational purposes.

Focusing only on the experimental spaces, the largest laboratory is the LSC Hall, combining a square dome measuring 22 m (width) by 22 m (length) by 8 m (height) on top, with a cylindrical pit having a diameter and depth of 20 m. The pit has a volume of 6,200 m$^3$ and serves as a multipurpose space for detectors. The second largest space, the AMoRE Hall, was excavated with dimensions of 21 m (width) by 21 m (length) by 16 m (height). Additionally, there are intermediate-sized spaces: one measuring 12 m (width) by 17 m (length) by 10 m (height), three measuring 8 m (width) by 15 m (length) by 8 m (height), and two measuring 7 m (width) by 25 m (length) by 7 m (height). The remaining spaces comprise sections with cross-sections of 5 m (width) by 5 m (height).

**4.2 Excavation**

The New Austrian Tunneling Method (NATM) was employed to ensure the safety and stability of the underground experimental spaces during rock excavation. The excavation of NATM tunnels involves a cyclical process consisting of five stages. Before the tunnel excavation, ground surveys were conducted, followed by dynamite blasting. Subsequently, excavation face reinforcement was carried out using rock bolts and shotcrete. Precise measurements of rock deformation were then conducted. If no deformation was detected, this process was repeated to proceed with the excavation.

Reinforcement methods (Type 1 to 5) were selected based on the Rock Mass Rating (RMR) grades (grade 1 to grade 5) [1]. Rock bolts and standard shotcrete, as detailed in Table 1, were applied, and stability measurements were taken before and after the blast. The table provides information on the radioactive isotope of the shotcrete components. The rock conditions at the excavation face were relatively advantageous, exhibiting RMR grades of 2 and 3. In addition to ensuring the safety of the experimental spaces, we took a conservative approach in applying reinforcement methods to protect the people working in the tunnel. Across the entire underground excavation tunnel, reinforcement methods corresponding to RMR grade 2 were applied in 38% of the area, and RMR grade 3 reinforcement methods were applied in 62%. All sections of the experimental spaces were reinforced using RMR grade 3 methods. Most of the rock in the experimental area is composed of limestone, with some sections showing signs of metamorphic and quartz veins.

The rock samples collected during the blasting process and the primary aggregates used for reinforcement were analyzed for the isotopic concentrations of U, Th, and K using ICP-MS, and the values are presented in Table 1.

The deepest point in the experimental tunnel is the AMoRE Hall, with a depth from the surface to the hall floor at 1,029 m (~2,500 m w.e.). The LSC Hall reaches a depth of 980 m, and most other spaces in the experimental area have depths exceeding 1,000 m (Figure 3). The total volume of excavation is approximately 65,000 m$^3$, and all excavation work was completed by August 2022.

**5 Infrastructures**

**5.1 Safety**

We conducted fire spread simulations using the 'ANSYS Fluent' [2] program to analyze the tunnel's structure, electrical systems, and ventilation facilities. Based on the results, we employed the 'Simulex' program [3] to devise effective evacuation strategies within the underground experimental space. Subsequently, we determined the refuge locations within the experimental area and installed a refuge from MINEARC Co., Ltd (Australia) [4] to accommodate up to 40 individuals during evacuation. In addition, we installed fire detection sensors such as smoke and flame detectors and high-definition cameras at regular intervals for early fire detection. It allows for effective environmental monitoring within the tunnel. Through these facilities, we aim to suppress the occurrence and spread of fires. As we operate the facilities, we will continuously work to enhance safety measures by addressing any weak points.

## 5.2 Electricity

Yemilab's underground facilities have a total power supply capacity of 1,600 kW (2,000 KVA) to support the power requirements for experiments. Approximately 200 kW is allocated for the operation of facilities such as lighting, ventilation, exhaust systems, etc., with the remaining 1,400 kW available for experiments. Critical facilities and primary experimental equipment are continuously supported by a 260 kW UPS (Uninterrupted Power Supply) unit and a surface emergency generator in preparation for power outages. In the event of a power outage, the surface emergency generator activates within a few seconds. This emergency generator can supply an instantaneous power of 360 kW until regular power is restored. Depending on the criticality of the facility, emergency power can be dynamically distributed as needed.

## 5.3 Ventilation

Due to its sealed structure, natural ventilation is impossible within Yemilab's underground spaces. However, the entrance of the Yemilab access tunnel is connected to the mine's shaft and the rampway, which is 6 km long to the surface. Under normal circumstances, ventilation in the mine operates in a natural circulation system, where surface air is drawn in through the vertical access tunnel, passes through the mine's rampway, and exits back to the surface.

Yemilab's ventilation system introduces air from the vertical access tunnel, utilizing a one-meter diameter duct to deliver 39,000 $m^3$/h of air into every corner of the Yemilab tunnel. Among this, 12,000 $m^3$ is supplied to the outdoor unit room where the outdoor units of the air conditioning systems located in the experimental area are gathered. The heat load of the outdoor unit room is assumed to be approximately 100 kW to maintain an indoor temperature below 40°C. The remaining 27,000 $m^3$ is supplied to the experimental area, providing about ten air exchanges daily for the entire Yemilab volume. Due to this ventilation system, the underground temperature can be maintained at 26°C, and the radon concentration remains below 50 Bq/$m^3$. However, during the summer months, usually June to September, when surface temperatures are high, the natural ventilation within the mine slows down significantly. As a result, the radon concentration within Yemilab rises to ~2,000 Bq/$m^3$. To reduce this level to below 150 Bq/$m^3$, a radonless air supply system is currently being installed to supply surface air directly into the underground. Once this facility is completed in 2024, the radon concentration within Yemilab during the summer will be maintained below 150 Bq/$m^3$. Furthermore, investigations are underway to improve the natural ventilation system of the mine tunnel to address the high radon concentrations during the summer.

## 5.4 Communication & Network

Anywhere within Yemilab's underground facilities, users can access all commercial mobile phone services available in South Korea. By November 2023, 1st step network installation for data transmission connecting the underground to the surface office was completed. The transmission speed is approximately 1 Gbps, enabling remote control of underground experimental equipment such as AMoRE-II and COSINE-200 from the surface office and facilitating the transmission of large volumes of experimental data. By 2025, there are plans to implement an automatic access control system for users of the underground laboratory utilizing the network facilities.

## 5.5 Groundwater

The groundwater within Yemilab's underground is discharged at approximately 4 tonnes daily. This groundwater is collected in a 90-tonne capacity reservoir at the end of the access tunnel beneath the entrance. This amount is sufficient to serve as domestic water for up to 40 individuals utilizing the underground, although it is not used for drinking. The collected groundwater in Yemilab is sent to the main reservoir in the mine via a pumping system. After combined with all other groundwater within the mine, it is discharged to the surface through drainage pipes installed in the shaft. This entire process is automated but can be operated manually if necessary.

## 5.6 Sustaining Cleanliness

Due to its proximity to mining facilities, Yemilab is inevitably affected by the high-concentration dust generated in the mine. To minimize dust contamination, we installed two main dustproof doors at the access tunnel, physically separating the experimental area, access tunnel, and mine tunnel. The floor of the experimental area is coated with epoxy, and each laboratory's wall is painted to block dust generated from shotcrete. Vehicles and equipment operate separately for each designated area. Vehicles requiring access to the experimental area go through cleaning before entering each section. Personnel entering also follow specific paths, changing shoes and clothing. The dust level in the experimental area shows a PM10 (≤10 μm) under about 10 μg/m$^3$ after the epoxy floor coating, which is lower than the Korean government's recommended limit of 50 μg/m$^3$ for office environments. After completing Y2L's relocation, strict management will reduce it to below 5 μg/m$^3$. Yemilab is specially equipped with an Radon Reduction System (RRS). It can provide air with a very low radon concentration level, typically below 100 mBq/m$^3$ at a rate of 50 m$^3$/h, suitable for experimental spaces requiring low radon levels. Furthermore, the RRS air meets approximately Class 1,000 cleanroom standard. An additional RRS, with a capacity of 250 m$^3$/h, will be installed in 2025.

## 6 Radiation Background and Measurements

There were several measurement systems to understand the radiation environment of the Yemilab and monitor the seasonal variation. First, RAD7, the most popular radon detector, has been used to measure the radon levels in the Yemilab. The RAD7 was installed at AMoRE Hall, refuge, HPGe, and IBS tunnel end in Fig. 3 to see the difference through the Yemilab tunnel. Fortunately, all the values were consistent in each place because of the proper ventilation in the Yemilab. However, as mentioned in section 5.3, the limited ventilation caused a high radon level of up to 2000 Bq/m$^3$ during the summer. There are additional systems to improve this high radon level, like Rn-less air supply and permanent air circulation system through the Handuk iron mine. To confirm the performances of the additional systems, ionization-type radon detectors are installed at AMoRE Hall, HPGe room, and the end of the IBS tunnel. We continuously provide the radon levels from the radon detectors on the online monitoring system.

The second is the external gamma rays and neutrons. We have collected rock samples through the Yemilab tunnels to measure the amount of elements using the ICP-MS and radioactivity using the HPGe detector. The compositions of the light elements in the rock samples are especially essential to understanding the origin of the neutrons [5]. Several $^3$He detectors with different thicknesses of the moderators, already used to measure the neutron flux at Y2L [6], have been installed at AMoRE Hall, IBS tunnels end, and LSC PIT in Fig. 3 and measured for two months at each place. The ongoing analysis will be compared with the simulation using the element compositions of the rock samples.

The last is the muon flux at the Yemilab. The AMoRE-II muon veto system is installed, and the bottom array of the system can be used to measure the muon flux at the Yemilab. This bottom array of the muon veto system comprises twenty-two plastic scintillators, each with dimensions of 1680 x 310 x 61 mm$^3$, and accumulated data for a week. A preliminary muon flux was measured to be 1.0 x 10$^{-7}$ $\mu$/cm$^2$·s, four times less than the muon flux at Y2L, 3.8 x 10$^{-7}$ $\mu$/cm$^2$·s [7]. The measured value is consistent with an expectation value of 8.2 x 10$^{-8}$ $\mu$/cm$^2$·s using a GEANT4 simulation [8]. To profile the muons for each cavern precisely, we plan to install a thick plastic (or liquid) scintillator for long-time measurements at different caverns and provide the values to the experiments, which want to avoid the muon background.

## 7    Experimental Programs at Yemilab

AMoRE experiment is to search the neutrino-less double beta decay of Mo-100 isotopes. It will use about 160 kg of Li$_2$$^{100}$MoO$_4$ crystals cooled at about 10 mK and coupled with low-temperature sensors. It has 25 cm Pb and 70 cm polyethylene and water shielding. The estimated background level at ROI of the signal induced by muons at Yemilab is below 10$^{-5}$ counts/keV/kg/year. AMoRE requires a Class 100 cleanroom for crystal assembly on site and low humidity below 1%RH due to the hygroscopicity of the crystals. The dilution refrigerator will be installed at the center of the shielding in 2024 [9].

The COSINE-100 experiment will be upgraded to COSINE-100U after reassembling the crystals used in the COSINE-100 experiment. The copper box containing about 2 tonnes of liquid scintillator will be reused, and the shielding will be remade. The whole setup will be cooled to about -30°C in a refrigerator. The background is dominated by internal components, and radon-reduced air will be flown continuously to the copper box. COSINE-200 experiment with new NaI crystals containing a lower internal background will begin after the crystals are successfully grown [10].

A large liquid scintillator detector will be installed at the LSC pit. The pit will be laminated, and the laminated cylinder will be 19.5 meters in diameter and 22 meters in height. A detector of about two kilo-tonne liquid scintillator capable of separating Cherenkov light from scintillator light is planned for solar neutrino studies and sterile neutrino searches.

A general-purpose dilution refrigerator and cryostat will be installed for low-mass dark matter searches and R&D for detector tests. In addition to the dark matter and neutrino physics, a microgravity experiment, an experiment for anomaly searches with Na-22 sources, and other rare phenomena searches will be conducted.

## 8    Conclusion

While Yemilab construction was completed in October 2022, an expansion of essential facilities to support experiments is needed. After relocating Y2L facilities to Yemilab in 2024, the focus will be on the operation of Yemilab, with a particular emphasis on the operation of experiments such as AMoRE-II and COSINE-200. Step-by-step improvements will be implemented as needed. Additionally, we aim to contribute to various international underground experiment facilities worldwide by sharing information about resources and environments.

**Figures**

Figure 1. The red dot indicates Yemilab's location in Gangwon province, South Korea.

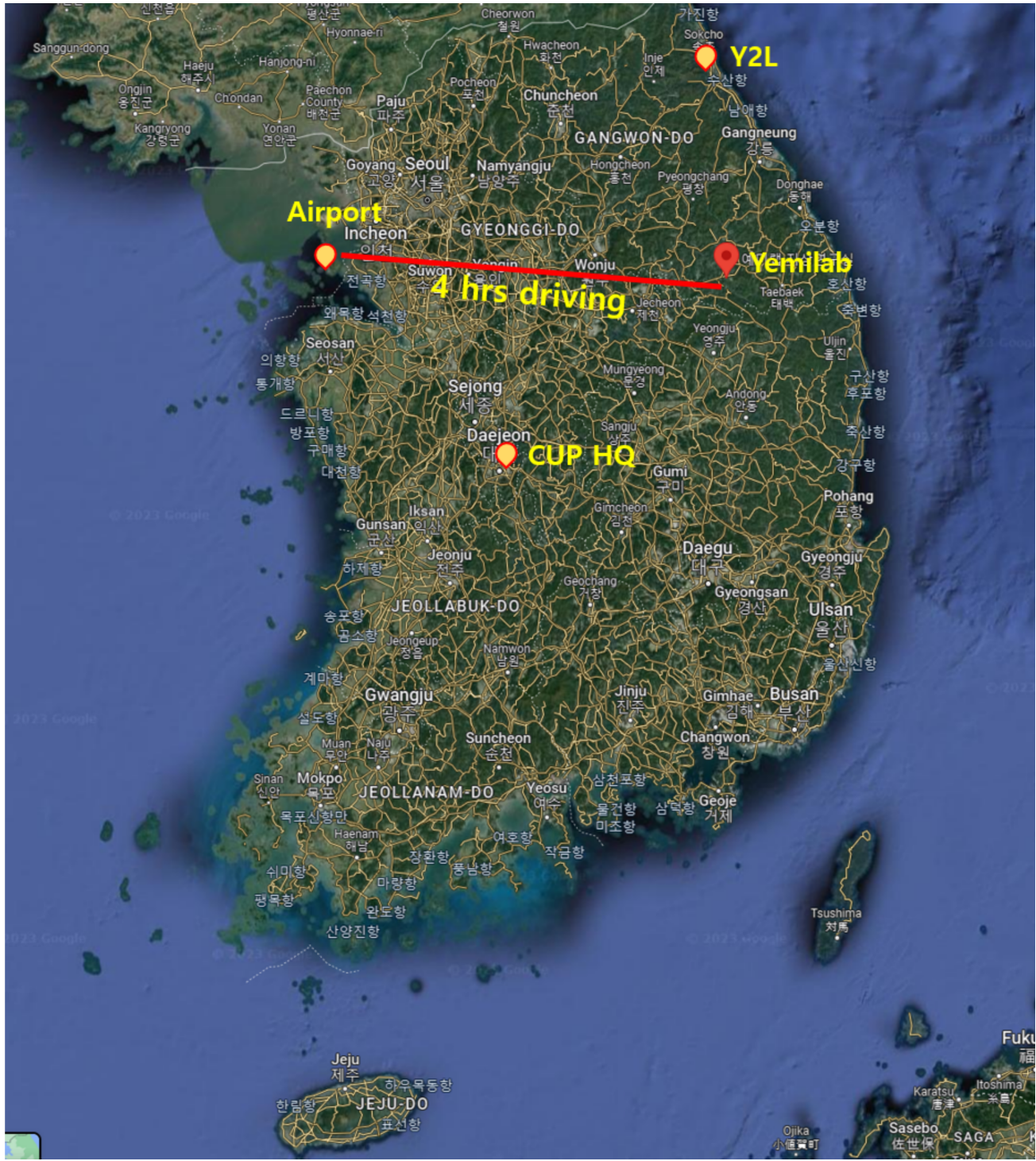

Figure 2. A view of the experimental zone. A photo was taken from the main hallway towards the end of the tunnel.

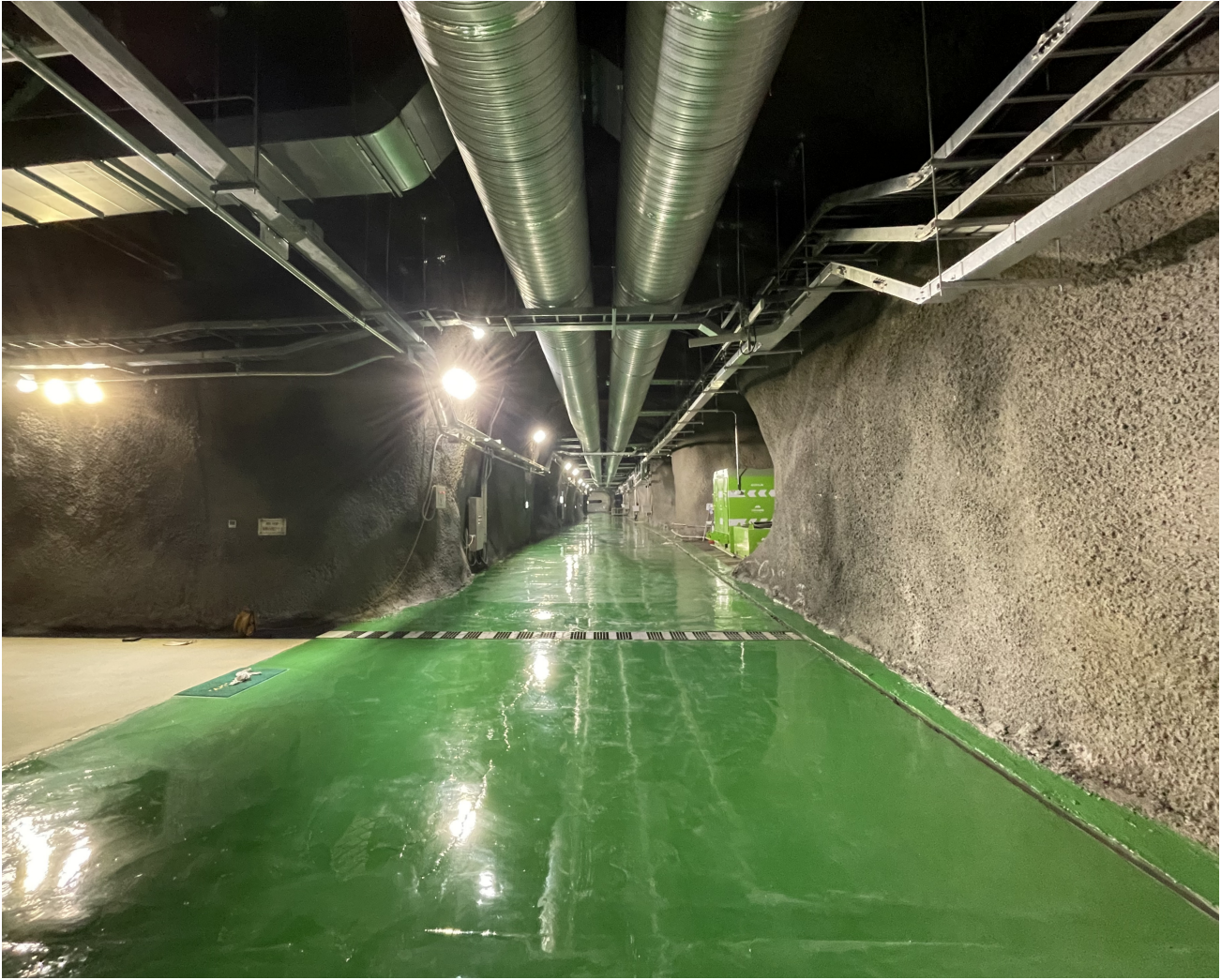

Figure 3. The structure of the underground experimental zone of Yemilab. The deepest point is the AMoRE hall floor with 1,029 m of overburden. The distance from the left end (Electricity) to the right end (KNU) is approximately 260 m.

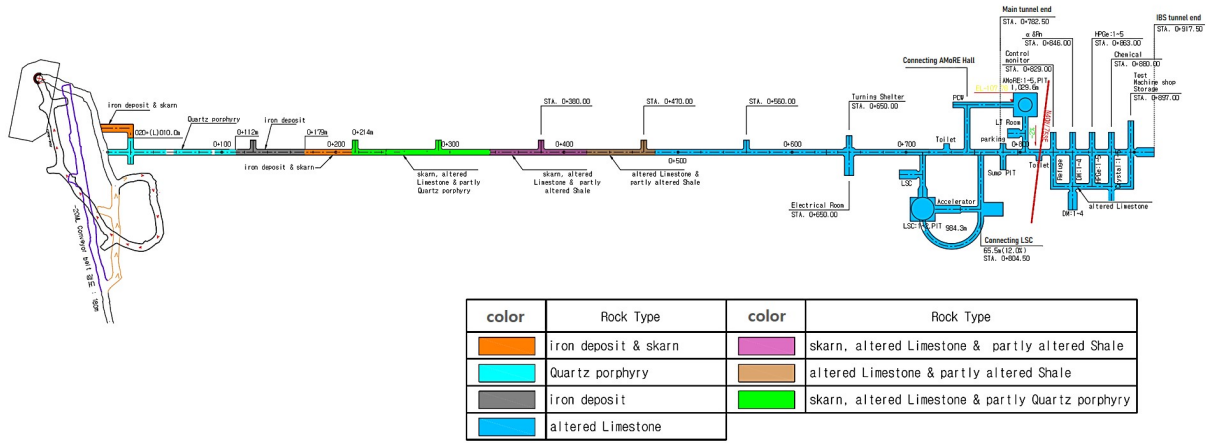

**Tables**

Table 1. The isotopic composition measurements and results of bedrock and aggregates of Yemilab.

| Rock | U (mg/kg) | Th (mg/kg) | K (mg/kg) | Remark |
|---|---|---|---|---|
| Access tunnel | 1.21 | 9.62 | 24,267 | @ sump pit |
| AMoRE-Hall | 0.84 | 3.27 | 11,800 | |
| LSC pit | 1.58 | 7.15 | 24,600 | |
| Shotcrete | U (mg/kg) | Th (mg/kg) | K (mg/kg) | Mixture Ratio (%) |
| Sand | 1.98 | 13.05 | 27,384 | 49 |
| Gravel | 0.72 | 2.17 | 1,768 | 27 |
| Cement | 2.10 | 5.24 | 6,977 | 22 |
| Steel fiber | 0.22 | 0.39 | 610 | 2 |
| Concrete for experimental room's floor | U (mg/kg) | Th (mg/kg) | K (mg/kg) | Mixture Ratio (%) |
| Sand | 0.50 | 2.05 | 4,300 | 50 |
| Gravel | 0.82 | 1.41 | 4,900 | 28 |
| Cement | 2.10 | 5.24 | 6,977 | 22 |